\documentclass[10pt,letterpaper]{article}
\usepackage[
  paper=letterpaper,
  top=0.85in,
  left=1in,      
  right=1in,     
  footskip=0.75in
]{geometry}

\usepackage{amsmath,amssymb}
\usepackage{changepage}
\usepackage{textcomp,marvosym}
\usepackage{cite}
\usepackage{nameref,hyperref}
\usepackage[right]{lineno}
\usepackage[nopatch=eqnum]{microtype}
\DisableLigatures[f]{encoding = *, family = * }

\usepackage[table]{xcolor}
\usepackage{float}
\usepackage{booktabs}
\usepackage{tabularx}
\usepackage{multirow}
\usepackage{array} 
\usepackage{graphicx} 
\usepackage[export]{adjustbox}
\usepackage{caption}
\usepackage[list=true]{subcaption}
\usepackage{listings}

\lstdefinelanguage{Julia}{
  morekeywords={function, end, if, else, elseif, for, in, while, return, using, import, struct, mutable, abstract, type, where},
  sensitive=true,
  morecomment=[l]\#,
  morestring=[b]",
}

\usepackage{array}
\newcolumntype{+}{!{\vrule width 2pt}}
\newlength\savedwidth

\usepackage[aboveskip=1pt,labelfont=bf,labelsep=period,justification=raggedright,singlelinecheck=off]{caption}

\makeatletter
\renewcommand{\@biblabel}[1]{\quad#1.}
\makeatother

\usepackage{lastpage,fancyhdr,graphicx}
\usepackage{epstopdf}
\fancyheadoffset[L]{2.25in}
\fancyfootoffset[L]{2.25in}

\begin{document}
\vspace*{0.2in}

\begin{flushleft}
{\Large
\textbf\newline{Exploring epidemic control policies using nonlinear programming and mathematical models} 
}
\newline
\\
Sandra Montes-Olivas\textsuperscript{1*},
Adam J. Kucharski\textsuperscript{1},
Michael B. Gravenor \textsuperscript{2},
Simon D.W. Frost\textsuperscript{1,3}
\\
\bigskip
\textbf{1} Department of Infectious Disease Epidemiology and Dynamics, London School of Hygiene \& Tropical Medicine, London, UK
\\
\textbf{2} Department of Health Data Science, Faculty of Medicine, Health \& Life Science, Swansea University, Swansea, United Kingdom
\\
\textbf{3} Microsoft Discovery and Quantum, Microsoft, Redmond, United States of America
\bigskip

* sandra.montes-olivas@lshtm.ac.uk

\end{flushleft}

\section*{Abstract}
Optimal control theory in epidemiology has been used to establish the most effective intervention strategies for managing and mitigating the spread of infectious diseases while considering constraints and costs. Using Pontryagin’s Maximum Principle, indirect methods provide necessary optimality conditions by transforming the control problem into a two-point boundary value problem. However, these approaches are often sensitive to initial guesses and can be computationally challenging, especially when dealing with complex constraints. In contrast, direct methods, which discretise the optimal control problem into a nonlinear programming (NLP) formulation, hold potential for automation and could offer robust, adaptable solutions for real-time decision-making. However, despite their potential, the widespread adoption of these techniques has been limited. Several factors may contribute to this challenge, including restricted access to specialised software, a perception of high computational costs, or a general unfamiliarity with these methods. \\

This study investigates the feasibility, robustness, and potential of direct optimal control methods using nonlinear programming solvers on compartmental models described by ordinary differential equations to determine the best application of various interventions, including non-pharmaceutical interventions (NPIs) and vaccination strategies. Through case studies, we demonstrate the use of NLP solvers to determine the optimal application of interventions based on single objectives, such as minimising total infections, ``flattening the curve'', or reducing peak infection levels, as well as multi-objective optimisation to achieve the best combination of interventions.\\

While indirect methods provide useful theoretical insights, direct approaches may be a better fit for the fast-evolving challenges of real-world epidemiology. By integrating newly available data more
quickly, direct methods can enhance the ability to make informed and timely decisions for managing outbreaks effectively.

\section*{Author summary}
This study highlights the advantages of direct optimisation methods in epidemiological modelling when there is a need to identify effective strategies for disease control while balancing constraints. Through case studies, it examines the effort required to adapt compartmental models for optimisation, the time needed to obtain an optimal solution, and the performance of both open-source and licensed tools. 
The study begins by contrasting indirect and direct methods using a simple infection model. It then explores the application of a mathematical programming framework, JuMP, to optimise control strategies aimed at reducing infections, minimising intervention costs under constraints, and managing multiple interventions. Finally, the study compares the efficiency of different optimisation algorithms. The results indicate that direct methods, aided by readily available tools like JuMP and IPOPT, enable efficient, flexible, and interpretable modelling with minimal additional implementation effort. This research highlights how these techniques can support informed and timely decision-making in the early stages of an epidemic. 

\section*{Introduction}

The study of infectious disease dynamics is crucial for the development and assessment of effective control strategies. Mathematical modelling has become an indispensable tool in this field, providing insights into transmission mechanisms and the potential impact of various interventions. Optimal control theory in epidemiology has been used to establish the most effective intervention strategies for managing and mitigating the spread of infectious diseases while considering constraints and costs \cite{sharomi_2017}. Optimal control can provide a framework to balance common control strategies, such as vaccination and isolation, according to an objective function which can be aimed at minimising infections or intervention costs, enabling decision-makers to weigh the benefits of disease control against potential economic and societal impacts.

The literature on infectious disease modelling using optimal control has grown considerably, particularly in response to the recent coronavirus disease 2019 (COVID-19) pandemic \cite{who_2022}. For instance, Godara et al. \cite{godara_2021} applied control theory to a Susceptible-Infected-Recovered (SIR) model to minimise the total number of infected individuals while considering the costs of mitigation efforts. Their findings underscored the importance of strict initial interventions, followed by a gradual relaxation as the epidemic progresses and herd immunity increases. Similarly, Britton and Leskela \cite{britton_2023} explored optimal non-pharmaceutical interventions (NPIs) within a SIR framework, emphasising that a single, intense lockdown of short duration, implemented at an optimal time, was the most effective strategy for reducing infections while managing cumulative intervention costs. Other studies have highlighted the practical aspects of different control strategies. For example, Miclo et al. \cite{miclo_2022} focused on suppression policies to prevent healthcare systems from becoming overwhelmed. They proposed a ``filling the box" strategy that involves intensifying suppression measures as infections approach ICU capacity and relaxing them as pressure on healthcare systems decreases. Meanwhile, Asamoah et al. \cite{asamoah_2021} developed an optimal control model and performed a cost-effectiveness analysis of different combinations of various NPIs, including social distancing, personal hygiene, and disinfection of public spaces.

Beyond single-strain models, Arruda et al. \cite{arruda_2021} addressed reinfection and multiple viral strains using a multi-strain SEIR model, with a case study of the COVID-19 outbreak. Their study highlighted the need to consider the population's waning immunity and the emergence of new variants when implementing control strategies. Xia et al. \cite{xia_2024} proposed a geographically tailored optimal control strategy that underscored the importance of spatial control measures, such as border closures, to contain infectious disease outbreaks effectively. Finally, integrating medical and non-medical interventions, Smirnova \cite{smirnova_2024} suggested the need for control strategies that involve behavioural changes during the early stages of an outbreak and the importance of developing improved control tools, such as vaccines and therapeutics, to apply them as the efficacy of early interventions diminishes over time.  

Together, these studies showcase the versatility and impact of optimal control theory in guiding infectious disease management. They also point out the ongoing challenges in effectively applying these methods, particularly the need for approaches that are both adaptable to real-world complexities and practical for rapid implementation. Addressing these challenges requires a closer look at the methodologies used in optimal control.

Optimal control methods can be classified into two categories: indirect and direct \cite{caillau_2023}. Indirect methods rely on deriving first-order necessary conditions for optimality, typically using the Pontryagin’s Maximum Principle (PMP). The PMP states that if the control $u^*(t)$ and state $x^*(t)$ trajectories are optimal, then there exists a piecewise differentiable adjoint variable $\lambda(t)$ that satisfies conditions defined by the Hamiltonian of the problem \cite{lenhart_2007}. Indirect methods are known for their high numerical accuracy, but have significant challenges. Their reliance on precise initialisation makes convergence highly sensitive to the initial guess, and they often require detailed structural knowledge of the optimal solution, such as \textit{a priori} identifying bang-bang or singular arcs. In bang-bang controls, the optimal strategy switches instantaneously between its minimum and maximum allowable values. This results in a control function that is piecewise constant and characterised by sharp transitions, and potentially difficult to identify. The challenge from a singular arc occurs when the control does not operate at either extreme. Instead, it is determined by higher-order conditions, which require solving additional equations to characterise the control trajectory within that specific interval. These complexities introduce extra analytical and computational challenges. As a result, indirect methods can become intricate and time-consuming, particularly for complex or poorly understood models where the solution structure is not immediately apparent.

In contrast, direct methods adopt a more intuitive approach by discretising the model in time, transforming the continuous optimal control problem into a nonlinear optimisation problem (NLP)\cite{betts_2010}. If the discretised NLP is differentiable, it can be addressed using gradient-based optimisation solvers such as the Interior Point Optimizer (IPOPT) solver\cite{wachter_2006} or sequential quadratic programming (SQP) solvers\cite{nocedal_2006}. Non-differentiable problems can alternatively be solved by non-gradient-based methods, including metaheuristic algorithms such as Genetic Algorithms\cite{michalewicz_2013} and Simulated Annealing\cite{kirkpatrick_1983, aarts_2005}. 

Recent advances have expanded the optimisation landscape for mechanistic epidemiological models through the integration of deep learning techniques. Colas et al. \cite{colas_2021} proposed a toolbox in Python aimed at bridging epidemiological models and reinforcement learning, specifically leveraging Q-Learning combined with deep neural networks (DQN) \cite{mnih_2015} alongside evolutionary optimisation methods such as NSGA-II \cite{deb_2002}. Furthermore, Yin et al. \cite{yin_2023} proposed a hybrid optimisation approach integrating traditional optimisation frameworks with deep-learning algorithms, employing both first- and second-order optimisers such as Adam \cite{kingma_2014, zhang_2018} and L-BFGS \cite{nocedal_1999, moritz_2016}. Although these methods show potential, their computational demands can create challenges and may limit their applicability in time-critical scenarios.

In contrast, established NLP methods are often characterised by rapid convergence to locally optimal solutions thus minimising computational demands. These methods can handle complex system dynamics, multiple constraints, and continuously evolving models. Therefore, they may prove advantageous for facilitating real-time intervention planning and decision-making. However, despite their practical advantages, they have yet to be widely adopted in the context of epidemiological modelling. Factors such as the lack of access to specialised software, perceived high computational costs, and unfamiliarity with these techniques may hinder their widespread use. Additionally, practical challenges, like adapting complex epidemiological models to fit into an optimal control framework, can create barriers to broader adoption. 

This study aims to investigate the feasibility, robustness, and potential of automated optimal control methods in epidemiological modelling. Initial explorations have been conducted to assess the resources required for implementing these methods and to evaluate their reliability. This research seeks to demonstrate the use of direct methods of optimal control and their capabilities by tackling challenges such as computational complexity and adaptability. By demonstrating strategies to overcome these obstacles, we aim to illustrate the applicability of direct methods in optimal control. We hope this will encourage further studies to adopt and refine these approaches, ultimately contributing to more effective and timely decision-making in the management of infectious diseases.

\section*{Materials and methods}

\subsection*{Optimisation algorithms}

Epidemiological control problems frequently present nonlinear dynamics, and are constrained by the available resources \cite{hernandez-hargas_2022}. Based on the existing literature, many epidemiological models that employ optimal control theory rely on the PMP to derive necessary conditions, which are then solved using numerical methods. Among these, the forward-backward sweep method \cite{lenhart_2007} is one of the most commonly employed approaches for simulations. In contrast, direct optimisation methods are widely used in several engineering disciplines to find optimal solutions to complex problems \cite{betts_2010}. In these fields, a direct optimisation approach formulates the entire optimisation problem (states, controls, and constraints) in a single framework, as a NLP for instance, which is then solved numerically. The NLP solver methods can be grouped as sequential or simultaneous. Sequential methods, like SQP, involve breaking down the problem into a sequence of subproblems that are updated from previous iterations. On the other hand, simultaneous methods solve the optimisation problem as a whole by discretising both the state and control profiles in time, which may be better suited for large scale problems \cite{biegler_2007}. IPOPT is a commonly used open-source solver that is based on an interior-point approach to solve large-scale problems by iteratively improving a candidate solution from within the feasible region \cite{wachter_2006}. Here, IPOPT was chosen for its proven ability to efficiently solve large-scale nonlinear programming problems, and its wide availability across multiple programming environments.  

JuMP (Julia Mathematical Programming) is an open-source algebraic modelling language embedded in Julia \cite{dunning_2017, lubin2023}. Similar to other commercially available optimisation software such as AMPL (A Mathematical Programming Language) \cite{amplbook_2003} and GAMS (General Algebraic Modeling System) \cite{gams_2009}, JuMP enables users to express optimisation problems in a form similar to their original mathematical representation, which is then translated into the form expected by the solver. JuMP provides an interface to various solvers and supports automatic differentiation, making it highly efficient for large-scale optimisation problems. The main components of a NLP formulation in JuMP are as follows; a solver; a list of variables; constraints that define the feasible region for the solution; and the objective function used to minimise or maximise specific outcomes, such as the number of infections or costs. 

All models presented in this study were discretised using a basic Euler method, a straightforward and commonly used approach for approximating solutions. This discretisation helps maintain consistency in the behaviour of the model and transforms the continuous system into a form suitable for solving the NLP. The IPOPT solver was used to solve the NLP due to its open-source nature and broad support across the optimisation languages used in this study.

\subsection*{Model formulation}

To demonstrate our methodology, we begin with a simple exponential‐growth infection model in which a time‐dependent control intervention seeks to reduce the number of infected individuals. This example provides a setting to compare the two methodologies: the indirect method, in which the Pontryagin’s Maximum Principle is applied to derive optimality conditions, and the direct method, in which the problem is discretised and solved as a nonlinear programming problem. 

Next, we present the application of JuMP with IPOPT across four case scenarios. The first three scenarios evaluate single control intervention strategies on a modified SIR model aimed at minimising total infections in an epidemic through lockdown measures, 'flattening the curve,' or vaccination efforts. The final scenario evaluates a more complex compartmental model of dengue transmission formulated by Asamoah et al. \cite{asamoah_dengue_2021}, optimising the combination of four control strategies.

All case scenarios were solved on a MacBook Pro with an Apple M3 Pro chip and 18GB of memory. The computational framework employed JuMP (v1.24), IPOPT (v3.14.17), AMPL (v20241203), amplpy(v0.14.0), rAMPL (v2.0.13.0.20241012), AmplNLWriter (v1.2.3) and Pyomo (v6.8.2). Code illustrating these scenarios can be found at \url{http://github.com/EpiRecipes/EpiPolicies}.

\section*{Results}

\subsection*{Comparison of Indirect and Direct methodologies}

We consider a simple mechanistic model that tracks the number of infected individuals $I(t)$ over time, assuming a nearly constant susceptible population $S \approx N$, where $N$ denotes the total population size. The dynamics of infection are described by:

\begin{equation}
\frac{dI}{dt} = \left( \beta (1 - \upsilon(t)) N - \gamma \right) I,
\end{equation}

where $\beta$ is the transmission rate, $\gamma$ is the recovery rate, and $\upsilon(t) \in [0, \upsilon_{\max}]$ represents a time-dependent control strategy (i.e., varying social distancing measures or lockdown) aimed at reducing transmission.

The optimal control policy $\upsilon(t)$ that balances disease mitigation with the cost of intervention is defined by the following objective function:

\begin{equation}
J = \int_0^{T_f} \left[ A I(t) + B \, \upsilon(t)^2 \right] \, dt,
\end{equation}

where $A$ and $B$ are the infection and control effort weights, respectively.

First, PMP was used to construct the Hamiltonian:

\begin{equation}
\mathcal{H}(I, \upsilon, \lambda) = A I + B \upsilon^2 + \lambda \left[ (\beta(1 - \upsilon)N - \gamma) I \right],
\end{equation}

where $\lambda(t)$ is the adjoint variable associated with state $I(t)$. Then, the adjoint equation is given by:

\begin{equation}
\frac{d\lambda}{dt} = -\frac{\partial \mathcal{H}}{\partial I} = -A - \lambda \left( \beta(1 - \upsilon)N - \gamma \right),
\end{equation}

To obtain the optimal control, we solve the following condition:

\begin{equation}
\frac{\partial \mathcal{H}}{\partial \upsilon} = 2B \upsilon - \lambda \beta N I = 0,
\end{equation}

which yields the unconstrained optimal control:

\begin{equation}
\upsilon^*(t) = \frac{\lambda(t) \beta N I(t)}{2B}.
\end{equation}

Taking into account control bounds, the final expression for the optimal control is as follows:

\begin{equation}
\upsilon^*(t) = \min\left( \max\left( 0, \frac{\lambda(t) \beta N I(t)}{2B} \right), \upsilon_{\max} \right).
\end{equation}

Using these derivations, the optimal control problem was solved numerically using the forward-backward sweep method \cite{lenhart_2007}, which iteratively updates the state and adjoint equations until convergence. 

To compare with a direct approach, the optimisation problem was then formulated and solved using the algebraic modelling language JuMP \cite{lubin2023} with the the interior-point solver IPOPT \cite{wachter_2006}. This approach requires the user to specify a solver and define decision variables, constraints, and the objective function. JuMP then translates the algebraic representation of the problem into a standard form that the solver interprets to find an optimal solution.  

\begin{figure}[!h]
     \centering
     \begin{subfigure}[t]{0.45\textwidth}
         \centering
         \includegraphics[width=2.0in]{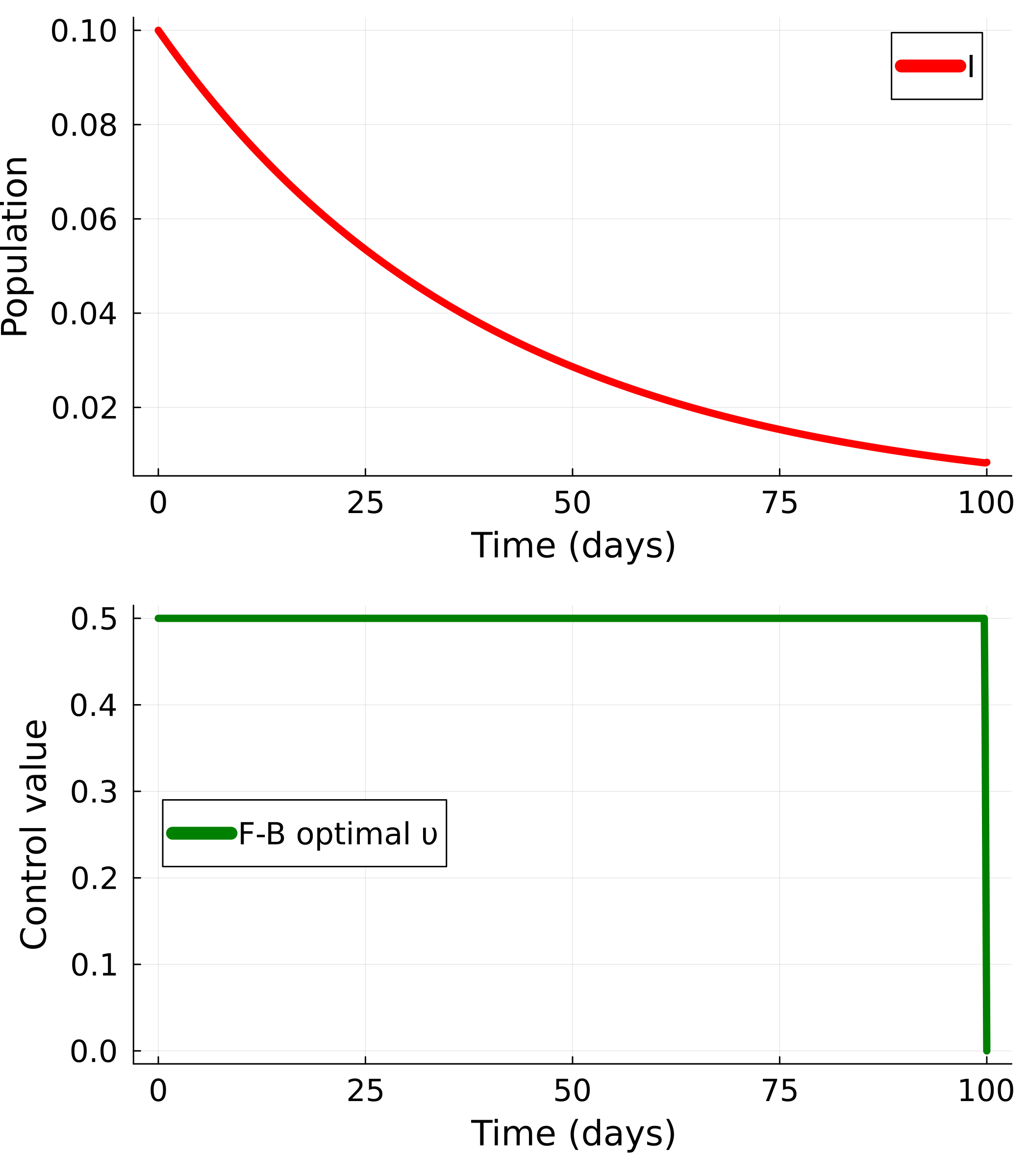}
         \captionsetup{justification=centering}
         \caption{Forward-backward sweep}
         \label{fig:FB_expinf}
         \hspace*{\fill}
     \end{subfigure}
     \begin{subfigure}[t]{0.45\textwidth}
         \centering
         \includegraphics[width=2.0in]{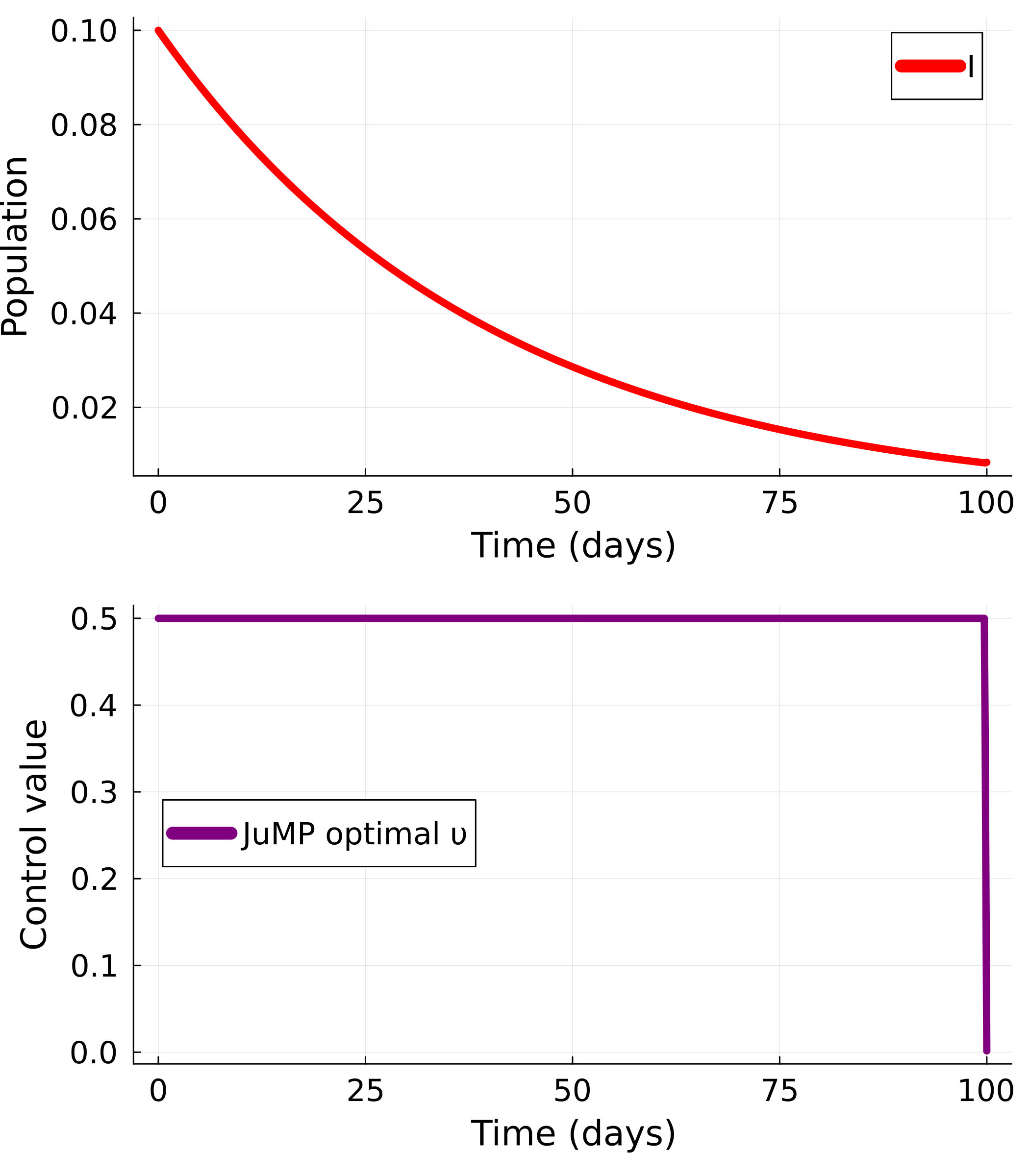}
         \captionsetup{justification=centering}
         \caption{JuMP + IPOPT}
         \label{fig:JuMP_expinf}
         \hspace*{\fill}
     \end{subfigure}
\caption{Optimal control solutions obtained using the (a) forward-backward sweep method and the (b) JuMP modeling framework with IPOPT}
\label{fig:ExpInf_opt_outputs}
\end{figure}

Code snippets illustrating the implementation of both methods are shown in Fig. \ref{fig:code_comparison}. The direct method not only results in a more concise block of code, but also eliminates the need for extensive preparatory analytical work, which can be considerable when applied to more complex models, compared to the simplified example presented here. Crucially, Fig. \ref{fig:ExpInf_opt_outputs} presents the results obtained from each approach, showing that both methods produced the same optimal solution.

\begin{figure}[!h]
  \centering

  \begin{minipage}[b]{0.48\textwidth}
    \centering
    \begin{lstlisting}[language=Julia, caption={Forward-backward sweep}, label={lst:forward_backward},
    mathescape=true]
        while test < 0 && sweep < max_iter
         sweep += 1
        
         I_old = copy(I)
         $\upsilon$_old = copy($\upsilon$)
         $\lambda$I_old = copy($\lambda$I)
        
         for k in 1:T
          infection = dt * $\beta$ * (1 - $\upsilon$[k]) * N * I[k]
          recovery  = dt * $\gamma$ * I[k]
          I[k+1] = I[k] + infection - recovery
         end
        
         $\lambda$I[T+1] = 0.0
         for k in T:-1:1
          $\lambda$I[k] = $\lambda$I[k+1] + dt * (-A + $\lambda$I[k+1] * ($\beta$ * (1 - $\upsilon$[k]) * N - $\gamma$))
         end
        
         temp = -$\lambda$I.*$\beta$.*N.*I./(2 .* B)
         $\upsilon$_new = clamp.(temp, 0.0, $\upsilon$_max)
         $\upsilon$ .= 0.5 .* ($\upsilon$_new .+ $\upsilon$_old)
        
         test = minimum([
          $\delta$ * sum(abs.($\upsilon$)) - sum(abs.($\upsilon$ .- $\upsilon$_old)),
          $\delta$ * sum(abs.(I)) - sum(abs.(I .- I_old)),
          $\delta$ * sum(abs.($\lambda$I)) - sum(abs.($\lambda$I .- $\lambda$I_old))])
         end
    \end{lstlisting}
  \end{minipage}%
  \hfill
  \begin{minipage}[b]{0.48\textwidth}
    \centering
    \begin{lstlisting}[language=Julia, caption={JuMP + IPOPT}, label={lst:jump_ipopt},
    mathescape=true]
using JuMP, Ipopt
model = Model(Ipopt.Optimizer)
set_optimizer_attribute(model, "max_iter", 1000)
@variable(model, 0 <= I[1:(T+1)] <= 1)
@variable(model, 0 <= $\upsilon$[1:(T+1)] <= $\upsilon$_max)
@expressions(model, begin
 infection[t in 1:T], (1 - $\upsilon$[t]) * $\beta$ * N * I[t] * dt  
 recovery[t in 1:T], $\gamma$ * dt * I[t] 
    end)
@constraints(model, begin
 I[1]==I0
 [t=1:T], I[t+1] == I[t] + infection[t] - recovery[t]
    end)
@objective(model, Min, 
            sum(dt * (A * I[t] + B * $\upsilon$[t]^2) for t in 1:T+1))
optimize!(model)
    \end{lstlisting}
  \end{minipage}

  \caption{Code snippets implementing the different approaches to solve the optimal control problem: forward-backward sweep (left) and JuMP with IPOPT solver (right).}
  \label{fig:code_comparison}
\end{figure}

\subsubsection*{Case scenario 1: Lockdown}

This scenario examines the optimal control of an SIR model through a lockdown intervention that reduces the infection rate, with the aim of minimising the total number of infected individuals. The key variables are as follows: $S$ represents the number of susceptible individuals, $I$ denotes the number of infected individuals, and $C$ indicates the total number of cases. The infection rate is modified according to a policy denoted as $\upsilon(t)$, where $0 \leq \upsilon(t) \leq \upsilon_{max} \leq 1$. This model is described by the following differential equations:

\begin{equation}
\begin{split}
\dfrac{\mathrm dS}{\mathrm dt} &= -\beta (1 - \upsilon(t)) S I, \\
\dfrac{\mathrm dI}{\mathrm dt} &= \beta (1 - \upsilon(t)) S I - \gamma I,\\ 
\dfrac{\mathrm dC}{\mathrm dt} &= \beta (1 - \upsilon(t)) S I\\
\end{split}
\label{eq:SIR_lockdown}
\end{equation}

where $\beta$ and $\gamma$ are the baseline transmission and recovery rates, respectively.

Britton and Leskela \cite{britton_2023} analytically demonstrated that the optimal policy under these conditions is a single ``bang-bang'' intervention, characterised by a sudden shift from no intervention to the maximum allowable level for a single period. The optimal control problem is defined as the policy that minimises the total number of cases (i.e., $C(\infty)$) while adhering to two main constraints: (a) the value of $\upsilon$ cannot exceed a specified maximum, $\upsilon_{max}$ , and (b) there is a cost associated with the policy, quantified as the integral of $\upsilon$ over time, which must remain within a certain limit i.e. $\int \upsilon(t) dt \leq \upsilon_{total}$. In the implementation, we calculate $C$ for a long time horizon to approximate $C(\infty)$.

\begin{table}[]
\centering
\caption{Values of parameters used in evaluations performed using the modified SIR models.}
\begin{tabularx}{0.8\textwidth}{ccl}
\toprule
\textbf{Parameter} & \textbf{Value} & \textbf{Definition} \\ 
\midrule
$\beta$               & 0.5     &  Transmission rate\\ 
$\gamma$             & 0.25    &  Recovery rate\\ 
S(0)               & 0.99    &  Initial fraction of suceptible population\\ 
I(0)               & 0.01    &  Initial fraction of infected population\\ 
$\upsilon_{max}$   & 0.5     &  Maximum policy value\\ 
$\upsilon_{total}$   & 10      &  Budget set for the control policy\\
$I_{max}$          & 0.5     &  Scenario 2: infected population threshold value\\ 
\bottomrule
\end{tabularx}

\label{tab:SIR_parameters}
\end{table}

\begin{figure}[!h]
     \centering
     \begin{subfigure}[t]{0.45\textwidth}
         \centering
         \includegraphics[width=2.0in]{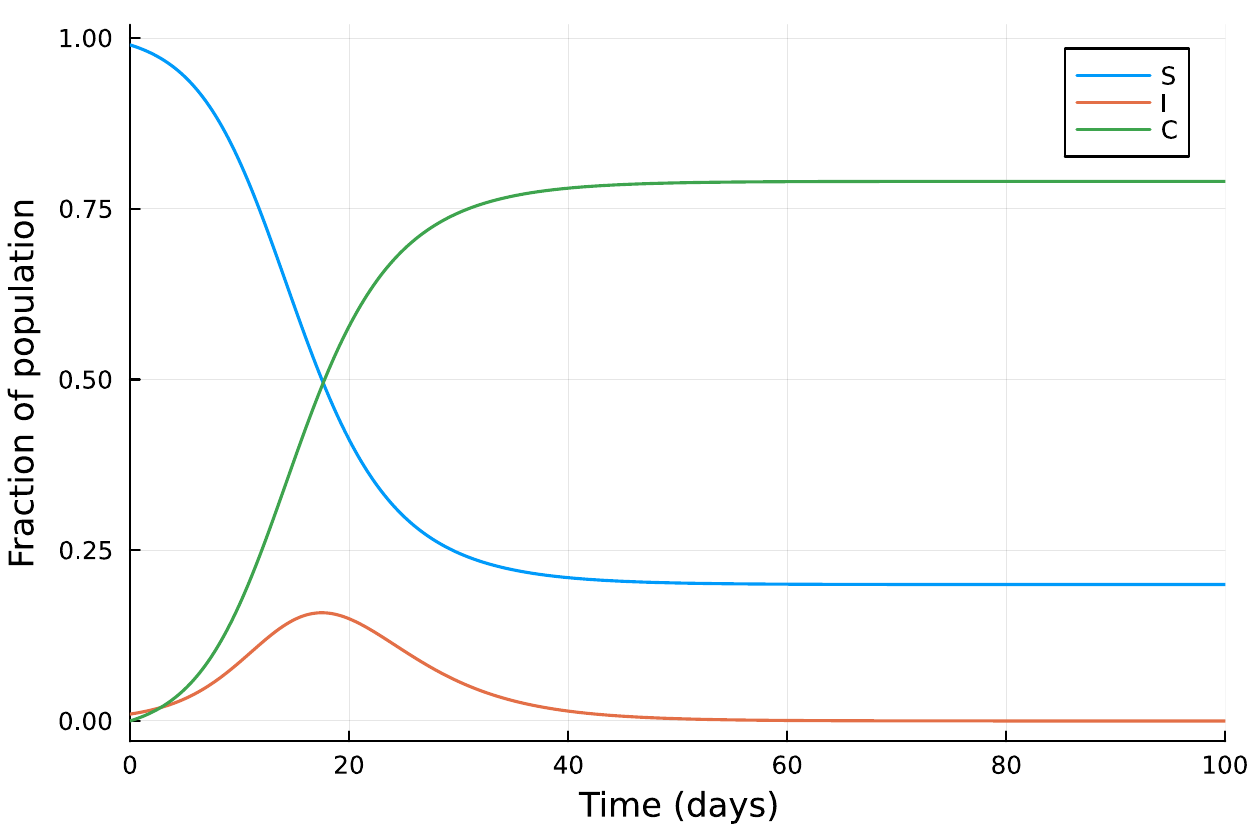}
         \captionsetup{justification=centering}
         \caption{SIR model baseline}
         \label{fig:SIR_baseline}
         \hspace*{\fill}
     \end{subfigure}
     \begin{subfigure}[t]{0.45\textwidth}
         \centering
         \includegraphics[width=2.0in]{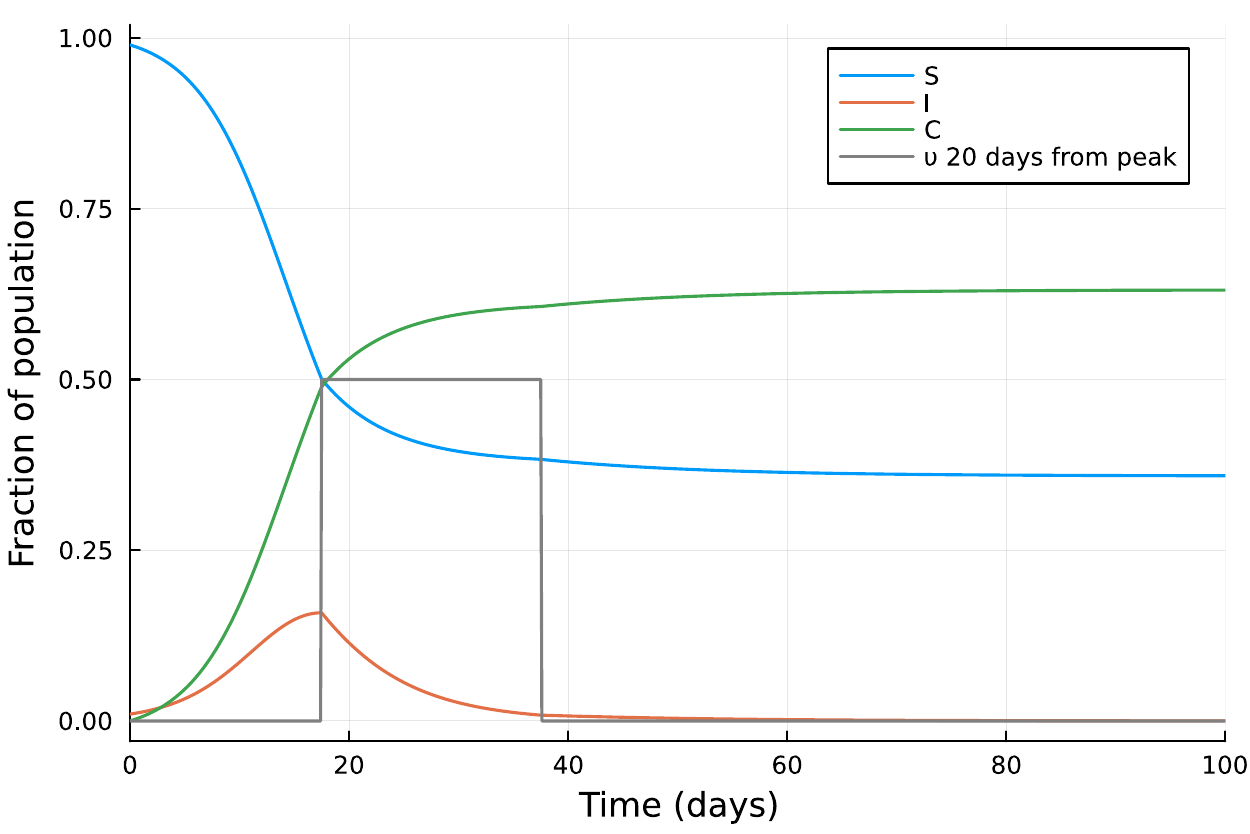}
         \captionsetup{justification=centering}
         \caption{Intervention start at peak}
         \label{fig:Interv_at_peak}
         \hspace*{\fill}
     \end{subfigure}
\caption{{\bf SIR baseline vs intervention starting at peak of infections.}
Comparison of SIR model outputs (a) baseline scenario without intervention and (b) with a lockdown intervention that lasts 20 simulated days, set at $\upsilon_{max}$ and applied at the peak of infections.}
\label{fig:SIR_Base_n_Peak}
\end{figure}

The parameter values used in the evaluations with the modified SIR models are listed in Table \ref{tab:SIR_parameters}, and they were assumed for the purpose of intervention optimisation only. These values were not derived from real-world data but were instead utilised to assess the performance of the optimisation algorithm within the model. We initially ran the model without the presence of intervention to establish baseline outputs, observing a cumulative incidence of approximately 79\% and the peak of infection occurring at around 17.5 days (see Fig.\ref{fig:SIR_baseline}).

This information was used to simulate the impact of a single lockdown intervention if it was initiated at the peak of infection cases and lasted for a fixed period of 20 days (set at the maximum policy value of 0.5) consistent with the duration suggested by Britton and Leskela \cite{britton_2023}. This scenario was chosen as representative of an intervention with a simple operational definition. Fig.\ref{fig:Interv_at_peak} shows that, in this case, the final cumulative incidence under intervention is 63\%.

\begin{figure}[!h]
     \centering
     \begin{subfigure}[t]{0.45\textwidth}
         \centering
         \includegraphics[width=2.0in]{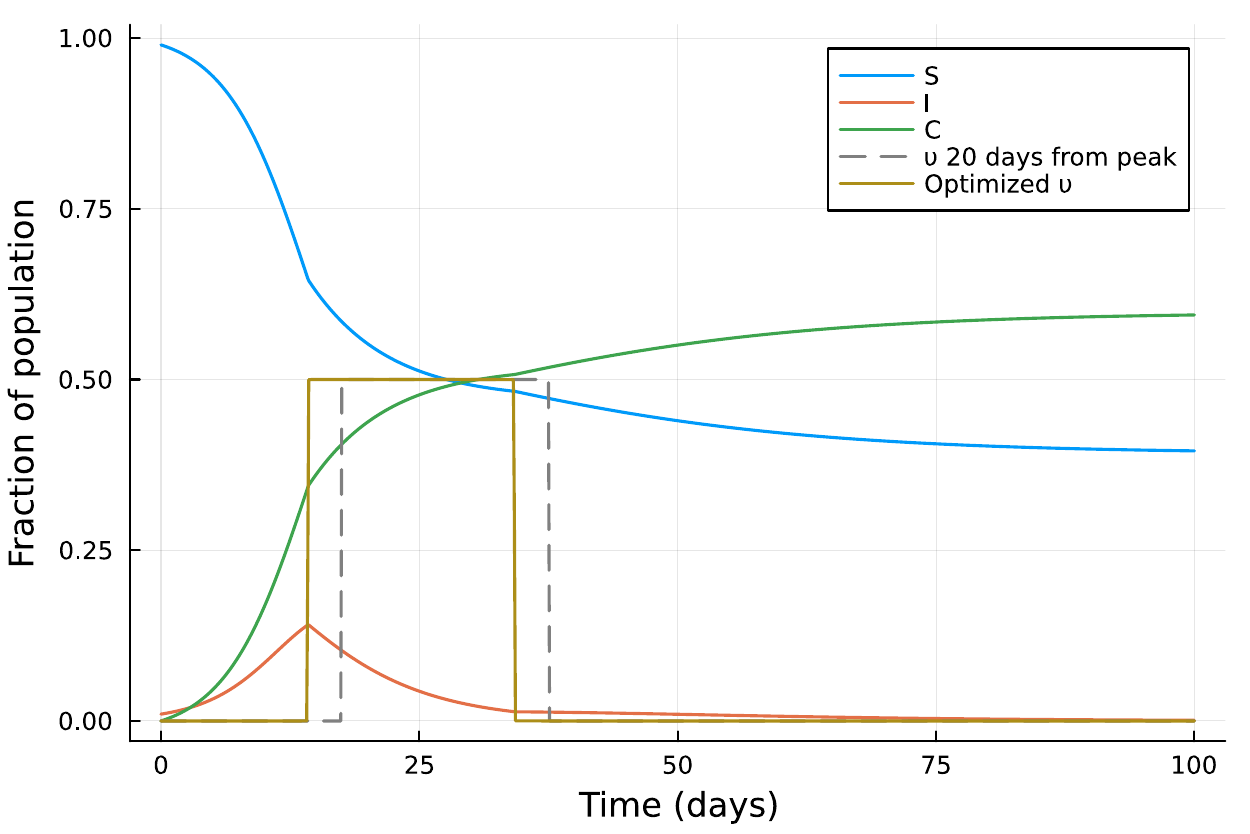}
         \captionsetup{justification=centering}
         \caption{Optimised lockdown}
         \label{fig:Lockdown_opt}
         \hspace*{\fill}
     \end{subfigure}
     \begin{subfigure}[t]{0.45\textwidth}
         \centering
         \includegraphics[width=2.0in]{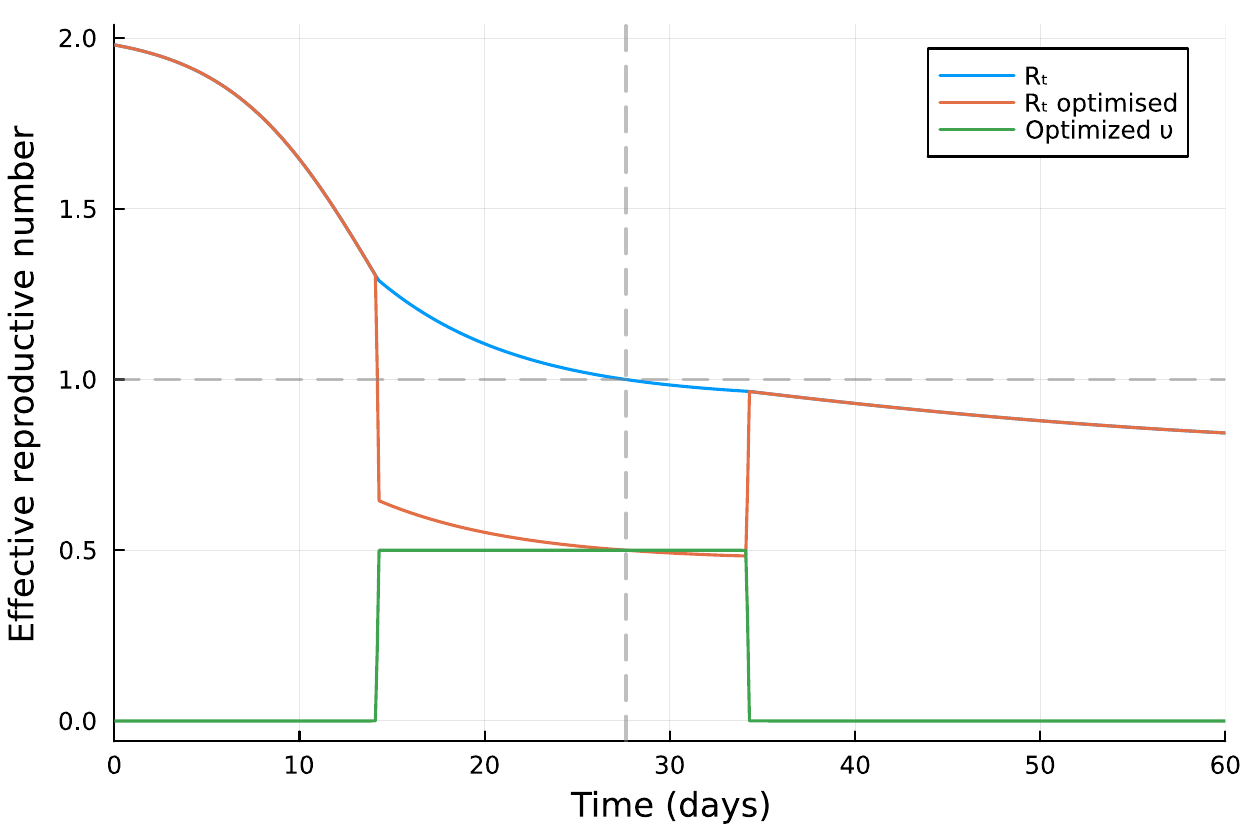}
         \captionsetup{justification=centering}
         \caption{$R_t$ at lockdown}
         \label{fig:Lockdown_opt_Rt}
         \hspace*{\fill}
     \end{subfigure}
     
     \bigskip
     
      \begin{subfigure}[t]{0.45\textwidth}
         \centering
         \includegraphics[width=2.0in]{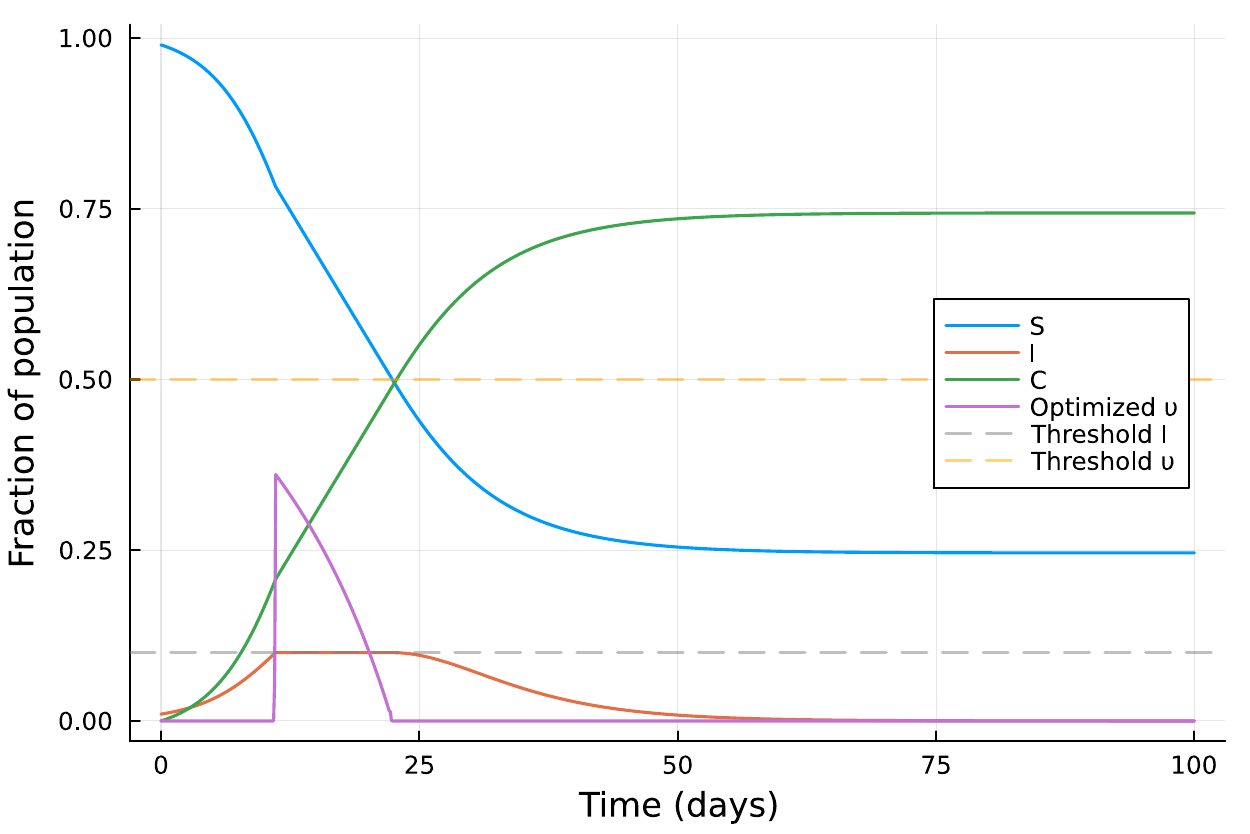}
         \captionsetup{justification=centering}
         \caption{Optimised FtC}
         \label{fig:Ftc_opt}
         \hspace*{\fill}
     \end{subfigure}
     \begin{subfigure}[t]{0.45\textwidth}
         \centering
         \includegraphics[width=2.0in]{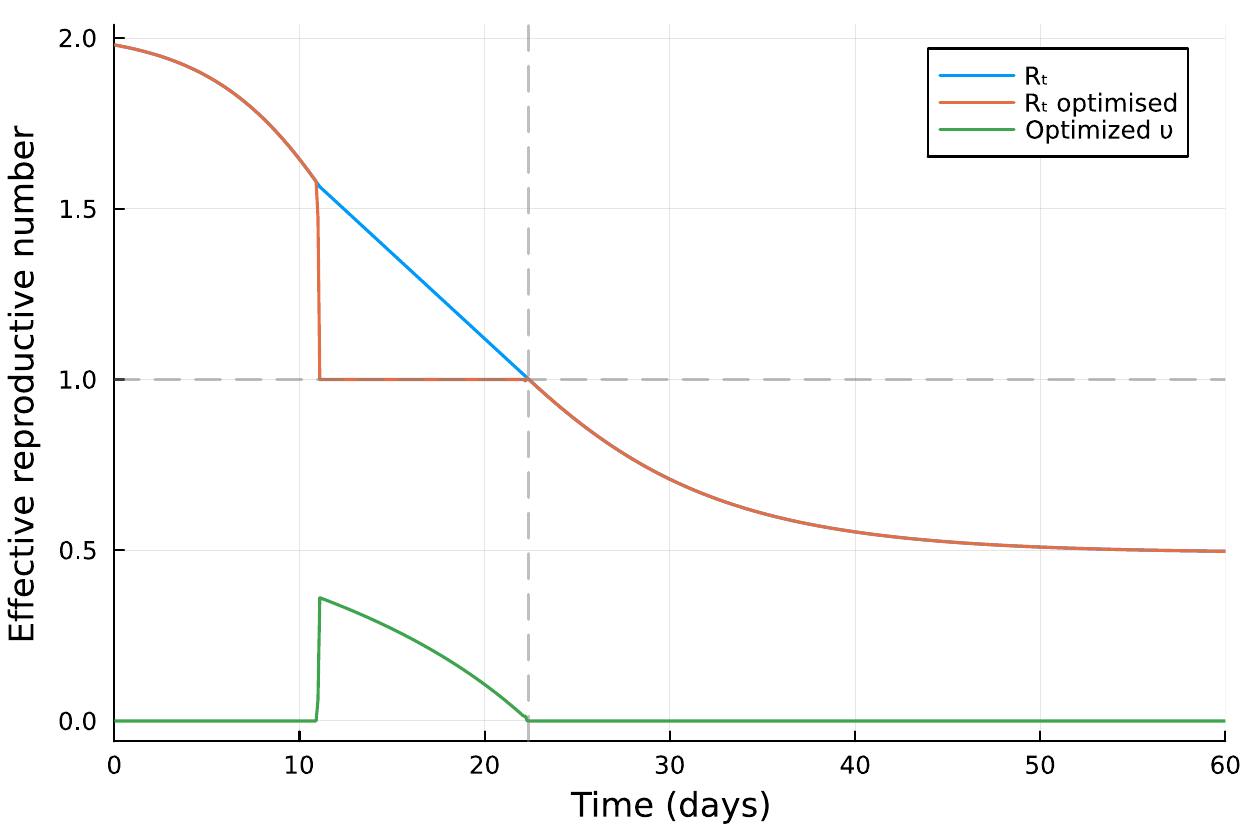}
         \captionsetup{justification=centering}
         \caption{$R_t$ at FtC}
         \label{fig:Ftc_opt_Rt}
         \hspace*{\fill}
     \end{subfigure}
\caption{{\bf SIR with optimised NPIs.}
Comparison of SIR model outputs (a) optimised lockdown scenario, (b) effective reproduction number during lockdown, (c) optimised ``flattening the curve'' (FtC) scenario, and (d) effective reproduction number during FtC intervention.}
\label{fig:SIR_Lock_n_Ftc}
\end{figure}

To investigate whether these results represent the optimal values for minimising cumulative incidence in this scenario, the NLP model was run. The results, shown in Fig.\ref{fig:Lockdown_opt}, indicate that the optimal start time obtained using JuMP is 14.3 days, which is earlier than the peak observed in the baseline non-intervention model (17.5 days). Furthermore, the optimal output suggests implementing a single lockdown lasting approximately 19.9 days, and the final cumulative incidence is 59\%.

\subsubsection*{Case scenario 2: ``Flattening the curve''}

This scenario employs the same system of equations \ref{eq:SIR_lockdown}, as in the lockdown scenario. In addition, in this case, the optimal intervention policy, represented as $\upsilon(t)$, balances the costs associated with intervention against the need to manage the spread of infection. This policy aims to ensure that the number of infected individuals, denoted as $I$, does not exceed a predetermined threshold, $I_{max}$. This approach is commonly referred to as ``flattening the curve'' (FtC), and it is easily implemented in the constraints included in the NLP. Both $\upsilon(t)$ and $I(t)$ are constrained to values between 0 and $\upsilon_max$ and $I_max$, respectively. The objective function in this scenario is to minimise the total cost of intervention instead of the cumulative incidence. 

The results shown in Fig.\ref{fig:Ftc_opt} suggest that the optimal policy following a FtC approach involves a single lockdown that increases in intensity rapidly at or shortly before the maximum tolerable infections ($I_{max}$). Once this threshold is reached the intensity of the lockdown is gradually reduced. Fig. \ref{fig:Ftc_opt_Rt} shows that the end of the lockdown coincides with the effective reproduction number, $R_t$, crossing the value of 1. This ensures that the population of infected individuals remains below the critical level.

\subsubsection*{Case scenario 3: Vaccination} 

Focusing on control through vaccination, the primary goal of the intervention is to effectively reduce the number of susceptible individuals within the population. This process is mathematically represented by the following set of equations \ref{eq:SIR_vacc}, where the main difference is the position of the control variable $\upsilon(t)$, which in this case represents vaccination at a per-capita rate.

\begin{equation}
\begin{split}
\dfrac{\mathrm dS}{\mathrm dt} &= -\beta S I - \upsilon(t) S, \\
\dfrac{\mathrm dI}{\mathrm dt} &= \beta S I - \gamma I,\\ 
\dfrac{\mathrm dC}{\mathrm dt} &= \beta S I\\
\end{split}
\label{eq:SIR_vacc}
\end{equation}

The optimal control problem is defined as the policy that minimises the total number of cases (i.e., the final size of the epidemic) while adhering to the following constraints: (a) the vaccination rate, $\upsilon$, cannot exceed a maximum value, indicating a limit on the rate of vaccination, and (b) there is a cost associated with the vaccination process, measured as the integral of $\upsilon(t)*S(t)$ over time, which cannot exceed a predetermined level. 

\begin{figure}[!h]
     \centering
     \includegraphics[width=0.5\textwidth]{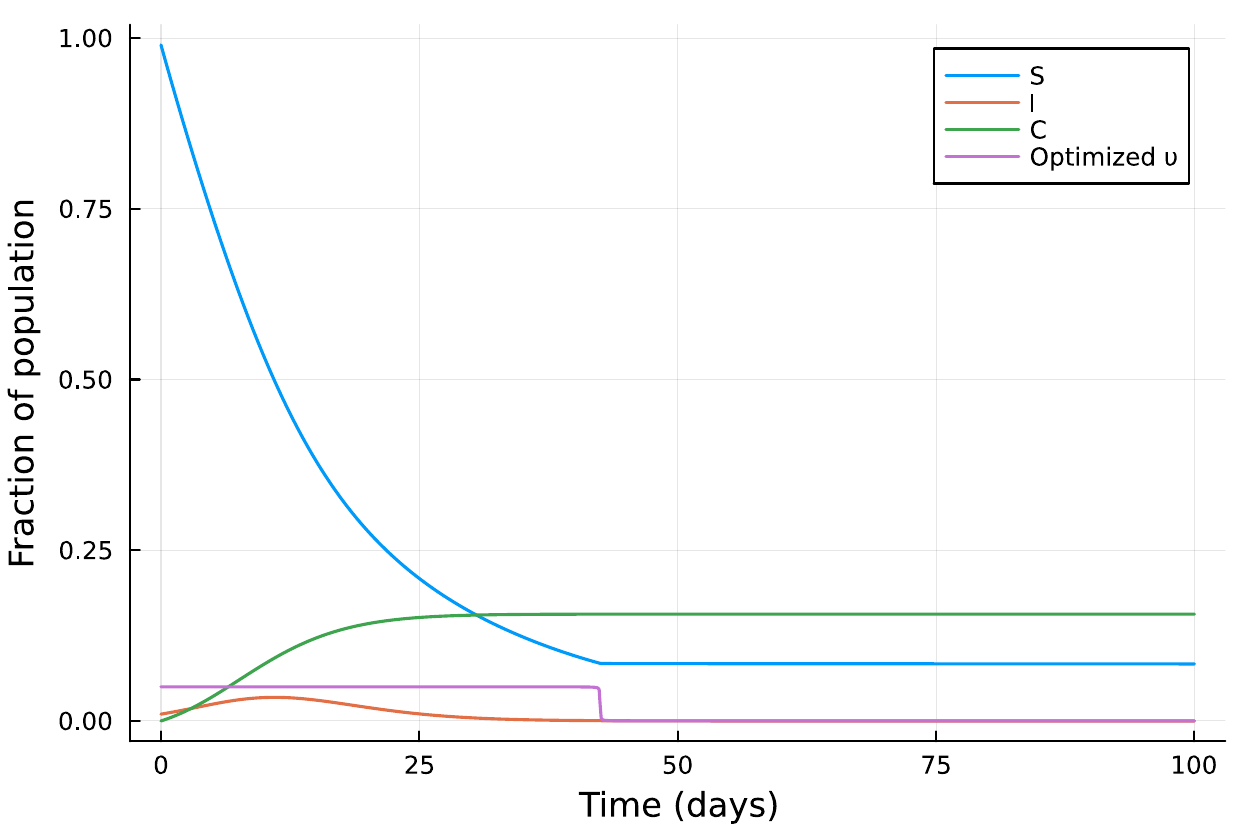}
\caption{{\bf SIR with optimised vaccination.}
Comparison of SIR model outputs (a) baseline scenario without intervention and (b) with a lockdown intervention that lasts 20 simulated days, set at $\upsilon\_max$ and applied at the peak of infections.}
\label{fig:Vacc_opt}
\end{figure}

The directly optimised results shown in Fig.\ref{fig:Vacc_opt} suggest that the optimal policy in this scenario is initiating vaccination immediately, and that maintaining a continuous administration at the maximum allowable rate is most effective. This intervention should be maintained until the allocated vaccine resources are depleted. This strategy reduces the susceptible population early in the epidemic, significantly limiting disease transmission and minimising the peak incidence of infections to $\sim 3\%$.

\subsubsection*{Case scenario 4: Multiple control strategies}

To assess the effectiveness of direct optimisation methods in identifying the optimal solution for a combination of multiple interventions, we applied the same methodology to a model proposed by Asamoah et al. \cite{asamoah_dengue_2021}, which provides a framework for evaluating the combination of different control interventions to control dengue transmission. This model considers two populations: human and female mosquito or vector populations. The human population is divided into five compartments: susceptible $S_h$, infected (symptomatic) $I_h$, carrier (asymptomatic) $I_hA$, partially immune $P$, and recovered $R_h$, while the mosquito population is formed by two compartments: susceptible $S_v$ and infected $I_v$. The model is described by the following equations.

\begin{equation}
\begin{split}
N_h(t) &= S_h(t) + I_h(t) + I_hA(t) + P(t) + R_h(t),\\
N_v(t) &= S_v(t) + I_v(t),\\
\lambda_h(t) &= \frac{(1 - u_1(t)) b \beta_1}{N_h(t)} I_v(t),\\
\lambda_h1(t) &= \frac{(1 - u_1(t)) b \beta_2}{N_h(t)} I_v(t),\\
\lambda_v(t) &= \frac{b \beta_3}{N_h(t)} (I_h(t) + I_hA(t)).
\end{split}
\label{eq:pop_dengue}
\end{equation}

\begin{equation}
\begin{split}
\frac{dS_h}{dt} &= \mu_h N_h - \lambda_h S_h - S_h u_2 - \mu_h S_h,\\
\frac{dI_h}{dt} &= \psi \lambda_h S_h + \omega \lambda_h1 P - (\mu_h + u_3 + \gamma_h) I_h,\\
\frac{dI_hA}{dt} &= (1 - \psi) \lambda_h S_h + (1 - \omega) \lambda_h1 P - (\mu_h + \gamma_h) I_hA,\\
\frac{dP}{dt} &= u_2 S_h + \rho u_3 I_h + \phi \gamma_h (I_h + I_hA) - \lambda_h1 P - \mu_h P,\\
\frac{dR_h}{dt} &= (1 - \rho) u_3 I_h + (1 - \phi) \gamma_h (I_h + I_hA) - \mu_h R_h,\\
\frac{dS_v}{dt} &= \mu_v N_v (1 - u_4) - \lambda_v S_v - \mu_v S_v - r_0 u_4 S_v,\\
\frac{dI_v}{dt} &= \lambda_v S_v - \mu_v I_v - r_0 u_4 I_v.
\end{split}
\label{eq:Dengue_model}
\end{equation}

The time-dependent controls in the model are: treated bednets ($u_1$), vaccination ($u_2$), treatment with prophylactics ($u_3$), and insecticides ($u_4$). All parameter definitions and values were obtained from \cite{asamoah_dengue_2021}, and restated in Table \ref{tab:dengue_pars} for reference.

\begin{table}[ht]
    \centering
    \caption{Parameter values and descriptions used on the dengue fever model.}
    \begin{tabular}{ccl}
        \toprule
        \textbf{Parameter} & \textbf{Value} & \textbf{Definition} \\
        \midrule
        $\beta_1$ & 0.75 & Transmission probability from $I_v$ to $S_h$\\
        $\beta_2$ & 0.375 & Transmission probability from $I_h$ to $S_v$\\
        $\beta_3$ & 0.75 & Transmission probability from $I_v$ to $P$\\
        $b$ & 0.5 & Avg. biting rate per mosquito per person \\
        $\rho$ & 0.01 & Proportion of treated individuals with partial immunity \\ 
        $\psi$ & 0.4 & Proportion of incidence rate from $S_h$ to $I_h$\\
        $\gamma_h$ & 0.3288 & Human's disease related death rate \\
        $\omega$ & 0.54 & Proportion of incidence rate from P to $I_h$\\
        $\mu_h$ & 0.0045 & Human's natural mortality and recruitment rate \\
        $\mu_v$ & 0.0323 & Vector's natural mortality and recruitment rate \\ 
        $\phi$ & 0.48 & Proportion of natural recovery \\
        $r_0$ & 0.005 & Enhanced death rate \\ 
        $u_1$ & \{0, 0.75\} & Treated bednet control strategy \\
        $u_2$ & \{0, 0.75\} & Vaccination control strategy \\
        $u_3$ & \{0, 0.75\} & Treatment with prophylactics control strategy \\
        $u_4$ & \{0, 0.75\} & Insecticides control strategy \\
        $C_{1,2,3}$ & 5 & Weights related to $I_h$, $I_hA$, $S_v$ and $I_v$ populations \\
        $D_1$ & 16.62 & Weight related to $u_1$ \\
        $D_2$ & 2.5 & Weight related to $u_2$ \\
        $D_3$ & 5 & Weight related to $u_3$ \\
        $D_4$ & 16.62 & Weight related to $u_4$ \\
        \bottomrule
    \end{tabular}
    \label{tab:dengue_pars}
\end{table}

Following the Asamoah et al. \cite{asamoah_dengue_2021} study, the objective was to minimise dengue incidence and the intervention cost according to the objective function shown in \ref{eq:dengue_objFun}, with weights relevant to diferent compartments ($C_{1...3}$), and those relative to each control variable ($D_{1...4}$).   

\begin{align}
\min \sum_{t=1}^{T} & \left( C_1 \cdot I_h[t] + C_2 \cdot I_hA[t] + C_3 \cdot \left( S_v[t] + I_v[t] \right) \right. \notag \\
& + \frac{1}{2} \left( D_1 \cdot u_1[t] + D_2 \cdot u_2[t] + D_3 \cdot u_3[t] + D_4 \cdot u_4[t] \right)^2 )
\label{eq:dengue_objFun}
\end{align}

Similar to the case scenarios presented previously, the system of equations was discretised using a simple Euler discretisation and optimised using JuMP and the IPOPT solver. All tested control combinations resulted in an optimal solution, and the number of iterations and the time taken to reach this solution are displayed in Table \ref{tab:dengue_JuMP}. These results indicate that the JuMP optimisation framework, in conjunction with the IPOPT solver, can return optimal solutions, even in more complex scenarios involving the optimisation of multiple control strategies. In this case, where a weighted sum is used to aggregate the contributions of each strategy to the objective function, the algorithms effectively converge to the optimal solution in less than 30 seconds for all combinations tried (see Table \ref{tab:dengue_JuMP}).

\begin{table}[ht]
    \centering
    \caption{Comparison of the number of iterations and the time taken to reach an optimal solution when combining control strategies using JuMP with the IPOPT solver.}
    \begin{tabular}{ccc}
        \toprule
        Interventions & No. Iterations & Time (s) \\
        \midrule
        $u_1 + u_4$ & 1201 & 24.252 \\
        $u_2 + u_3$ & 534 & 9.843 \\
        $u_1 + u_3 + u_4$ & 1432 & 27.629 \\
        $u_1 + u_2 + u_4$ & 1229 & 23.682 \\
        $u_1 + u_2 + u_3 + u_4$ & 1171 & 22.053 \\ 
        \bottomrule
    \end{tabular}
    \label{tab:dengue_JuMP}
\end{table}

\subsubsection*{Comparison between optimisation algorithms}

There could be concerns regarding the use of direct optimisation methods if these are limited to specific programming languages. To address this, we leveraged the accessibility of the IPOPT solver across various platforms. 

Table \ref{tab:opt_alg_comparison} shows the number of iterations and time each algorithm took to find an optimal solution for the lockdown model. The results indicate that the solver maintains a consistent number of iterations across platforms. However, the slight variation in performance time may be attributed to how each algorithm handles sparsity. In many optimisation problems, the matrices involved, such as Jacobians and Hessians, frequently exhibit significant sparsity, characterised by a predominance of zeros or empty values. This sparsity can enhance the efficiency of calculations performed by the solver and save memory. Some platforms can exploit sparsity more efficiently than others. As a result, even when employing the same solver and undergoing an equivalent number of iterations, a more effective management of sparsity can improve the overall performance. 

From the optimisation algorithms tested in this study, JuMP and Pyomo are open-source, while amplpy and rAMPL are official interfaces that allow the user to access the license-based optimisation language AMPL and its features from Python and R, respectively. On the other hand, AmplNLWriter is an open-source wrapper that also interacts with AMPL-enabled solvers but is maintained by the JuMP community.  

By applying the solver to the initial case scenario using multiple programming languages, we demonstrate that users can choose their preferred programming language while utilising either open-source or licensed optimisation platforms, depending on their needs.

\begin{table}[ht]
\caption{Comparison of different optimisation algorithms using IPOPT}
\begin{tabularx}{\textwidth}{p{2.5cm}p{3cm}p{1.7cm}p{2cm}p{2.5cm}}
\hline
\textbf{Programming} & \textbf{Optimisation} & \textbf{Iterations} & \textbf{Total time (s)} & \textbf{Lagrangian Hessian} \\ \hline
\multirow{2}{*}{Julia}  & JuMP                & 93 & 1.346 & 300 \\ 
                        & JuMP \& AmplNLWriter & 93 & 0.054 & 250 \\ \hline
\multirow{2}{*}{Python} & Pyomo               & 93 & 0.051 & 250 \\ 
                        & amplpy              & 93 & 0.058 & 250 \\ \hline
R                       & rAMPL               & 93 & 0.058 & 250 \\ \hline
AMPL                    & AMPL IDE            & 93 & 0.051 & 250 \\ \hline
\end{tabularx}
\label{tab:opt_alg_comparison}
\end{table}

\section*{Discussion}

Modelling underpinned a number of key policy decisions in the UK during the COVID-19 pandemic. As in the examples here, the timing, duration, and extent of lockdown were foremost among the policy options considered at several time points \cite{gravenor_2025}. In our experience, the modelling efforts were characterised by a distribution of scenarios generated across a grid search of assumed plausible epidemiological parameters, combined (usually indirectly) with great uncertainty associated with any costs considered. With tight time constraints involved in preparing modelling output for real-time decision-making, the ability to automate the search of policy options in a manner that can both save time and identify optimal choices offers considerable advantages, especially for complex models with long runtimes that are often used in informing decision-making.

The application of numerical simulations and control theory is commonplace in the analysis of complex models, especially in fields dealing with unpredictable dynamics and large-scale systems \cite{betts_2010}. While these simulations provide valuable insights into a wide range of complex systems, optimal control theory becomes particularly helpful when managing disease control, as it helps balance resource constraints and optimise treatment or control strategies. In addition, these strategies also aim to reduce the implementation costs associated with large-scale interventions, offering a double benefit of improving public health outcomes while minimising economic impact \cite{lenhart_2007, sharomi_2017}.

The continuous nature of many models, including those used in disease control, often necessitates discretisation for computational implementation. This study employed a simple Euler discretisation, a straightforward method for approximating continuous systems. While easy to implement, this method has limitations regarding accuracy and stability. To address these limitations, we also explored alternative discretisation techniques, such as using exponential approximations to model the transition probabilities in the modified SIR system. If we model the time between transitions (such as infections or recoveries) as an exponentially distributed random variable with rate $\lambda$, the probability of transition between states during the timestep can be approximated by $1-exp(-\lambda\cdot dt)$. This method may be more accurate for larger timesteps than the Euler approximation. However, smaller timesteps, although giving results closer to the continuous time system, resulted in the solver struggling to converge. 

In case scenario 1, the lockdown intervention was optimised; it is worth noting that other grid-search-like methods could have been employed. For instance, one approach could involve fixing the intervention length to 20 days and optimising only the start of the intervention. The supplementary \nameref{fig:S1_Fig} material, shows a comparison of the optimal time found using JuMP with the cumulative incidence obtained by running the model simulations for different intervention start times. However, the use of JuMP allowed us to confirm the optimal intervention frequency and duration.

Compared to case scenario 1, where the objective was to minimise the overall number of infections, case scenario 2 had a different focus, the goal being to keep the infected population below a certain threshold, which could be influenced by factors such as the current healthcare capacity. This led to a distinct optimal policy, which suggested a single lockdown period that started earlier than in scenario 1. However, unlike in scenario 1, the strength of the intervention gradually decreased over time until the effective reproduction number dropped below one, at which point disease transmission was effectively controlled. While this optimal policy provides a theoretical framework for managing disease transmission, there are significant challenges in translating these findings into actual intervention policies. Fine-tuning the intensity of the intervention over time, as suggested by the model, may not be feasible in real-world settings. Instead, a more practical approach may involve implementing a series of staged interventions with varying intensities, depending on the evolving situation. The impact of the intervention may be uncertain prior to its implementation, and if its efficacy is found to be lower than expected, it may require initiating the intervention well in advance before reaching the infected threshold. Furthermore, determining when to stop the intervention requires knowledge of the effective reproduction number in the absence of the intervention. This requires reliable estimates of $R_t$ and the intensity of the intervention. These uncertainties are in addition to the usual uncertainty in model structure and parameter values of the underlying model.

Similarly, case scenario 3 relies on the assumption that a vaccine is available at the start of an epidemic, and in such a case, the optimal policy would be to vaccinate early and at the maximum capacity available until the supply is exhausted, to reduce the susceptible fraction of population as much as possible. However, it is important to recognise that, in reality, even if a vaccine is developed quickly, there are several factors that can delay or limit its availability. The production capacity may be insufficient to meet demand, leading to delays in distribution. Additionally, the vaccine's efficacy may not be immediately known or consistent across all demographic groups. Variations in immune response based on age, health status, or pre-existing conditions could affect the overall success of the vaccination campaign.

Despite the obvious appeal of identifying optimal control strategies, there are challenges associated with solving optimisation problems. We highlight here that indirect methods are more mathematically intensive, and their successful application depends on an in-depth understanding of the system's dynamics and constraints. In contrast, the direct methods described do not require the explicit construction of the adjoint and control equations, making them more straightforward to define computationally. Nevertheless, they may come with limitations, such as potentially slower convergence or less precision in highly complex systems. Another issue may arise when the constraints of an NLP problem turn out to be infeasible, which is a common problem that the user should be prepared to address. In these cases,  the solver may fail to converge to a feasible solution, which may require revisiting and adjusting the model or constraints \cite{betts_2010}. Thus, it is essential to have a clear understanding of the system and the ability to adjust the model as necessary to avoid such pitfalls. Therefore, we do not advocate one approach as the standard; rather we want to provide the reader with examples of how optimisation languages such as JuMP can be a very useful tool to obtain fast optimal approximations or even corroborate the results obtained through indirect methods. Similarly, Silva et al. \cite{silva2013} applied both direct and indirect methods to develop optimal control strategies to minimise the cost of interventions for treating tuberculosis. They utilised AMPL, IPOPT, and PROPT MATLAB Optimal Control Software to compute and compare the numerical results with those obtained through an iterative method, a standard approach for solving systems of ODEs and updating controls in indirect methods.

As for solving NLP optimisation problems, using an interior point solver such as IPOPT has proven to be highly effective in this study and in others \cite{wachter_2006, biegler_2007, word_2012, silva2013, zhu_2023}. IPOPT can efficiently solve problems with a large number of inequalities and degrees of freedom, making it a good and flexible option for high-dimensional control problems. The algorithm's ability to scale and find feasible solutions within a reasonable time frame is an important advantage when tackling the optimisation challenges posed by complex control systems. Additionally, advancements in computing power have reduced the computational expense once associated with these methods, enabling faster and more accessible solutions. However, alternative tools such as AMPL, GAMS, and Pyomo also offer similar capabilities, providing users with flexibility in software selection based on preference and the specific requirements of their computational environment.

It is important to note that this study focused exclusively on differentiable epidemiological models, which allowed us to use gradient-based optimisation solvers. As a result, our findings do not directly address the effectiveness or applicability of non-gradient-based optimisation approaches. However, recent advancements have introduced differentiable frameworks for traditionally non-differentiable models, such as agent-based models (ABMs), which may expand the scope and use of gradient-based optimisation solvers. For instance, Chopra et al. \cite{chopra_2022}, developed GradABM, a scalable and differentiable design for ABMs that enables gradient-based learning through automatic differentiation.

\section*{Conclusion}

Overall, the NLP optimisation approach illustrated in this study performed very effectively for simple models, consistently delivering accurate results within seconds which showcase its potential as a practical tool for rapid analysis and decision-making in epidemiological modelling.

While JuMP and IPOPT have been used in previous studies to solve optimisation problems in the context of infectious disease modelling \cite{zhu_2023}, this study aims to showcase their ease of use and effectiveness to promote the wider adoption of direct methods for addressing optimisation challenges. By employing JuMP and IPOPT, this work illustrates how these tools can effectively deliver an initial solution. This approach can aid policymakers in making informed decisions during the early stages of an epidemic, based on the dynamics of the disease model.

Regarding ongoing challenges in public health economics, the uncertainty surrounding costs has been consistently highlighted in the evidence presented to the UK COVID Inquiry \cite{UKCovidInquiry_2024}. As we consider future developments, it is important to integrate cost considerations into our decision-making processes. Despite this need, it is likely that cost estimates will continue to be uncertain. In addition, there may be other considerations that would affect the optimal policy; for example, Zarebski et al. \cite{zarebski2025including} incorporated measures of equity into finding an optimal vaccination program in order to achieve an intervention that was both effective and ethical. Thus, there is a pressing need for efficient methodologies that facilitate the exploration of optimal strategies across a wide array of cost function assumptions. The methods described here offer an opportunity for real-time evaluation and the potential generation of rules of thumb for optimal approaches, based on extensive exploration of cost-benefit analysis under a wide range of disease dynamic scenarios, to support preparedness. 

\section*{Supporting information}

\paragraph*{S1 Fig.}
Final cumulative infected population values (C) obtained from applying a fixed intervention with a length of 20 days at different starting times. The minimal C value is highlighted in orange.\\
\centering
\includegraphics[width=0.5\linewidth]{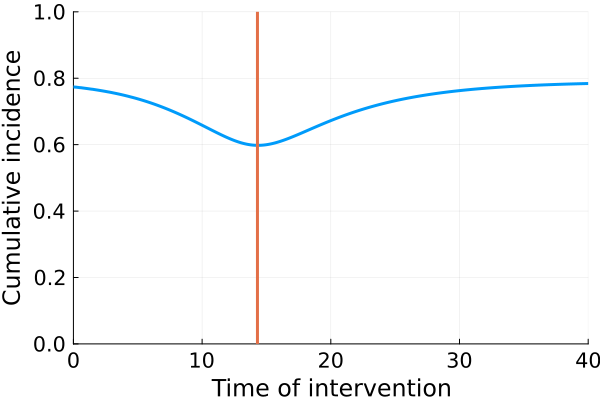}
\label{fig:S1_Fig}

\paragraph*{S2 Fig.}
Performance of an optimal intervention strategy to reduce the cummulative infections, as in case scenario 1, across various scenarios by sweeping through a range of transmission and recovery parameter values ([0.35,0.55] for $\beta$ and  [0.15,0.30] for $\gamma$). Rather than optimising for each scenario separately, the JuMP model creates a unified intervention profile that reduces the expected total infections across all.\\
\centering
\includegraphics[width=0.6\linewidth]{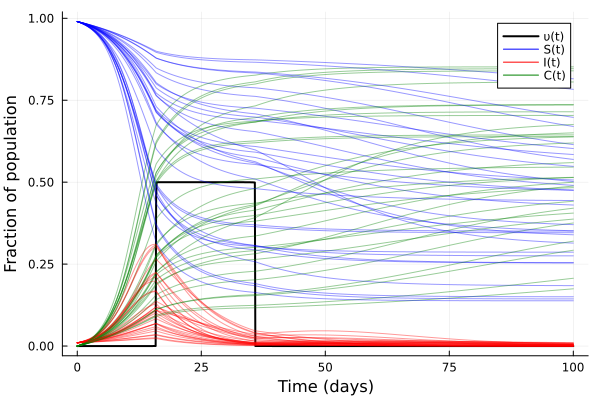}
\label{fig:S1_Fig}

\section*{Acknowledgments}
\raggedright
This work was supported by the Medical Research Council (MRC) under the grant ``Building an epidemiological modelling toolkit for epidemic preparedness'' (Grant Ref: MR/Z503939/1). The authors sincerely thank Dr. Joshua Asamoah, whose shared files and communication supported the testing of the methods presented in this study.

\nolinenumbers

\bibliography{references}

\end{document}